\documentclass{iau}
\usepackage{graphicx}
\usepackage{amsmath}
\newcommand{\apj}{ApJ}
\newcommand{\mnras}{MNRAS}
\newcommand{\aap}{A\&A}

\title[Young Stars Near the Sun] 
{Vertical Shear Instability in the Solar Nebula}

\author[M.-K. Lin \& A. N. Youdin]   
{Min-Kai Lin \and Andrew N. Youdin}

\affiliation{Department of Astronomy and Steward Observatory,\\ 
  University of Arizona, 933 North Cherry Avenue, Tucson, AZ 85721, USA 
  \\ email: {\tt minkailin@email.arizona.edu}, {\tt youdin@email.arizona.edu }}

\pubyear{2015}
\volume{314}  
\pagerange{119--126}
\jname{Young Stars \& Planets Near the Sun}
\editors{J. H. Kastner, B. Stelzer, \& S. A. Metchev, eds.}
\begin{document}

\maketitle

\begin{abstract}
  We quantify the thermodynamic requirement for the Vertical Shear
  Instability and evaluate its relevance to realistic protoplanetary 
  disks as a potential route to hydrodynamic turbulence. 
 
\end{abstract}

\firstsection 
\section{Introduction}
A purely hydrodynamic route to turbulence has important implications
for the evolution of cold protoplanetary disks and solids within
them. One such candidate is the Vertical Shear Instability (VSI) 
operating in disks where the orbital frequency $\Omega$ depends on the
height $z$ away from the disk midplane. While astrophysical disks generally
possess vertical shear, VSI also requires short thermal 
timescales. We determine a quantitative thermodynamic requirement for
the VSI, and apply our results to realistic disk conditions. We find 
the VSI can operate effectively at $5$---$50$AU in a typical protoplanetary disk,
with characteristic growth times of $30$ orbits.

\section{The need for rapid cooling for the VSI}
Astrophysical disks generally have vertical shear, $\p_z\Omega\neq
0$. This is a source of free energy, and may thus lead to
instability (\cite[Goldreich \& Schubert 1967]{goldreich67};
\cite[Urpin 2003]{urpin03}; \cite[Barker \& Latter 2015]{barker15}),
provided the disturbance vertical lengthscale is much larger than its
radial lengthscale. However, if the disk is stably stratified, as is typical
for irradiated protoplanetary disks, the VSI also requires rapid
cooling to overcome the stabilizing influence of vertical buoyancy
(\cite[Nelson et al. 2013]{nelson13}).  

More specifically, for a vertically isothermal disk with radial
temperature dependence $T\propto r^{q}$, the cooling or thermal
relaxation timescale $t_c$ must be sufficiently small for the VSI to
operate effectively:
\begin{align}\label{crit}
  t_c\Omega_\mathrm{K} < \frac{h|q|}{\gamma - 1}\equiv
  \beta_\mathrm{crit},  
\end{align}
where $\Omega_\mathrm{K}$ is the Keplerian frequency, $h$  is the disk
aspect ratio, and $\gamma$ is the adiabatic index (\cite[Lin \& Youdin
2015]{lin15}). 
Since protoplanetary disks are typically thin, $h\ll 1$,
Eq. \ref{crit} implies that $t_c \ll \Omega_\mathrm{K}$ is required
for the VSI. That is, the thermal timescale must be significantly
shorter than the dynamical timescale, which reflects the fact that
vertical shear is weakly destabilizing, while vertical buoyancy is
strongly stabilizing. 



\section{Applicability of the VSI in realistic protoplanetary disks} 
We solve the linear stability problem for axisymmetric
disturbances in a model of the Solar Nebula described in \cite[Chiang
\& Youdin (2010)]{chiang10}. We include in the energy equation a
realistic cooling function, based on dust opacity, which depends on
$z$ as well as the disturbance lengthscale.  

\begin{figure}
  \begin{center}
    \includegraphics[width=\linewidth]{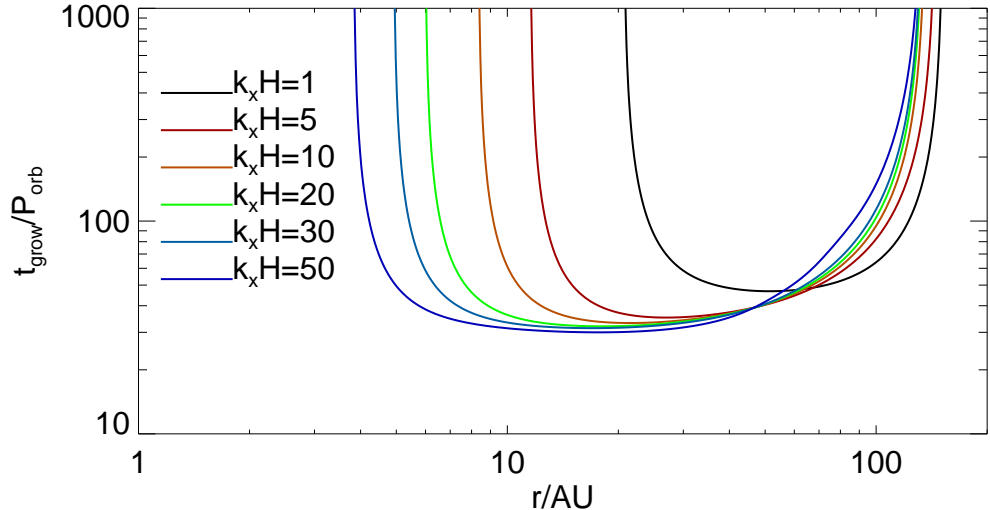} 
    \caption{Characteristic growth times of the Vertical Shear
      Instability in a typical protoplanetary disk model. Here, $k_x$
      is the wavenumber of the disturbance and $H$ is the disk
      scale-height. 
      \label{mmsn}}
  \end{center}
\end{figure}

In Fig. \ref{mmsn} we plot characteristic VSI growth times for a range
of radial perturbation wavenumbers. Growth times increase rapidly at $\lesssim5$AU
and $\gtrsim 100$AU as the cooling times violate Eq. \ref{crit} in
these regions. Thus, the VSI is most effective at intermediate
radii. Notice also the growth times are strongly dependent on
lengthscale toward the optically-thick inner disk, with smaller-scale
disturbances being able to operate further inwards. In the
optically-thin outer disk, cooling times are independent of scale and
all scales are stabilized beyond approximate the same radius.

\section{Conclusions}
Vertical shear is ubiquitous in astrophysical disks. Such
configurations are unstable if the cooling timescales are
significantly shorter than the dynamical timescale (Eq. \ref{crit}). 
This can be satisfied for small-scale disturbances at few 10s
of AU in typical protoplanetary disk 
models with realistic cooling. The VSI is thus dynamically important
in the outer disk, with potentially significantly implications for
transport and dust evolution.


\end{document}